\documentclass[pra,twocolumn,amsmath,amssymb,floatfix,reprint]{revtex4-1}

\pdfoutput=1
\usepackage{graphicx}
\usepackage{dcolumn}
\usepackage{bm}
\usepackage[usenames,dvipsnames]{xcolor}
\usepackage{mciteplus}
\usepackage{braket}
\usepackage{ulem}
\usepackage{url}

\renewcommand{\d}{\mathrm{d}}

\begin{document}

\title{Helicity, Topology and Kelvin Waves in reconnecting quantum knots}
\author{P.~Clark di Leoni$^1$, P.D.~Mininni$^1$, \& M.E.~Brachet$^2$}
\affiliation{$^1$Departamento de F\'\i sica, Facultad de Ciencias
    Exactas y Naturales, Universidad de Buenos Aires and IFIBA, CONICET,
    Ciudad Universitaria, 1428 Buenos Aires, Argentina.\\
                  $^2$Laboratoire de Physique Statistique de l'Ecole Normale
    Sup\'erieure associ\'e au CNRS et aux Universit\'es Paris 6 et 7, 24
    Rue Lhomond, 75237 Paris Cedex 05, France.}
\date{\today}

\begin{abstract}
    Helicity is a topological invariant that measures the linkage and
    knottedness of lines, tubes and ribbons. As such, it has found
    myriads of applications in astrophysics and solar physics, in
    fluid dynamics, in atmospheric sciences, and in biology. In
    quantum flows, where topology-changing reconnection events are a 
    staple, helicity appears as a key quantity to study. However, the
    usual definition of helicity is not well posed in quantum
    vortices, and its computation based on counting links and
    crossings of vortex lines can be downright impossible to apply in
    complex and turbulent scenarios. We present a new definition of
    helicity which overcomes these problems. With it, we show that
    only certain reconnection events conserve helicity. In other
    cases helicity can change abruptly during reconnection. 
    Furthermore, we show that these events can also excite Kelvin
    waves, which slowly deplete helicity as they interact nonlinearly,
    thus linking the theory of vortex knots with observations of
    quantum turbulence.
\end{abstract}
\maketitle


Helicity plays an important role in the dynamics of many fluid flows. It
is linked to the growth of large scale magnetic fields in astrophysics
and solar physics \cite{Brandenburg05}, the formation of supercell
convective storms in meteorology \cite{Rasmussen98}, the decay rate of
stratified turbulence \cite{Rorai13}, and the formation of large
structures in rotating and stratified flows \cite{Marino13} among other
problems. Helicity is a measure of the knottedness of field lines, which
is conserved under appropriate conditions, and as such it has been
called a ``topological invariant'' of many flows \cite{Moffatt69}. These
ideas \cite{Moffatt92,Dennis05,Ricca08,Moffatt14} have found
applications in areas beyond fluid dynamics, such as DNA biology
\cite{Vologodskii98}, optics \cite{Dennis10} and electromagnetism
\cite{Kedia13}.

Although helicity is perfectly conserved in barotropic ideal fluids, in
real fluids \cite{Hussain11,Kleckner13} and in superfluids
\cite{Bewley08,Zuccher12} vortex reconnection events, which alter the
topology of the flow, can take place. It is unclear how well helicity is
preserved under reconnection. As a few examples, experiments of vortex
knots in water have shown that center line helicity remains constant
throughout reconnection events \cite{Scheeler14}, while theoretical
arguments indicate that writhe (one component of the helicity) should be
conserved in anti-parallel reconnection events \cite{Laing15}, a fact
later confirmed in numerical simulations of a few specific quantum
vortex knots \cite{Zuccher15}. However, numerical studies of
Burgers-type vortices indicate that helicity is not conserved
\cite{Kimura14}. While experiments studying helicity in quantum flows
have not been done yet, the recent experimental creation of quantum
knots in a Bose-Einstein condensate in the laboratory \cite{Hall16} is
a significant step in that direction.

Recently, quantum flows have been used as a testbed for many of these
ideas \cite{Scheeler14,Zuccher15}, as vorticity in a quantum flow is
concentrated along vortex lines with quantized circulation, and as
these vortex lines can reconnect without dissipation. However, the
lack of a fluid-like definition of helicity for a quantum flow
requires complex topological measurements of the linking and
knottedness of vortex lines \cite{Scheeler14}, or artificial filtering
of the fields \cite{Zuccher15} to prevent spurious values of helicity
resulting from the singularity near quantum vortices. Moreover,
helicity in quantum flows has an interest {\it per se}, as
reconnection events in superfluid turbulence can excite Kelvin waves
\cite{Fonda14}. These are helical perturbations that travel along the
vortex lines first predicted for classical vortices by Lord Kelvin
(see \cite{Thomson80}). Kelvin waves are believed to be responsible
for the generation of an energy cascade \cite{Kozik04,Lvov10} leading
to Kelvin wave turbulence \cite{Clark15a}. Possible links between
helicity and the development of Kelvin wave turbulence have remain
obscure as a result of the difficulties involved in the measurement of
both helicity and Kelvin waves.

Here we study the time evolution of helicity and its link with Kelvin
waves in numerical simulations of the Gross-Pitaevskii equation (GPE).
The GPE models superfluids and Bose-Einstein condensates (BECs) (for
which the creation of quantum knots has been recently demonstrated in the
laboratory \cite{Hall16}) near zero temperature. We present a
fluid-like regularized definition for the helicity, which solves the
problem arising from the singularities in the velocity and
the vorticity produced by the topological defects of the quantum
flow, and links quantum knots with helicity as measured in fluid
dynamics. Then, we study the time evolution of helicity in multiple
linked rings and knots. We show that only some reconnection events
conserve helicity, while in others helicity changes abruptly during
reconnection, and later decays slowly towards a new constant value. We
link this depletion of helicity to the excitation of a Kelvin wave
cascade and the radiation of phonons. Finally, we illustrate how the
regularized helicity can be successfully used to quantify the helicity
in complicated and fully turbulent situations, such as a flow with
initial large-scale helicity, and where computation of helicity by
topological means would be impractical.

\begin{figure}
    \centering
    \includegraphics[width=8.5cm]{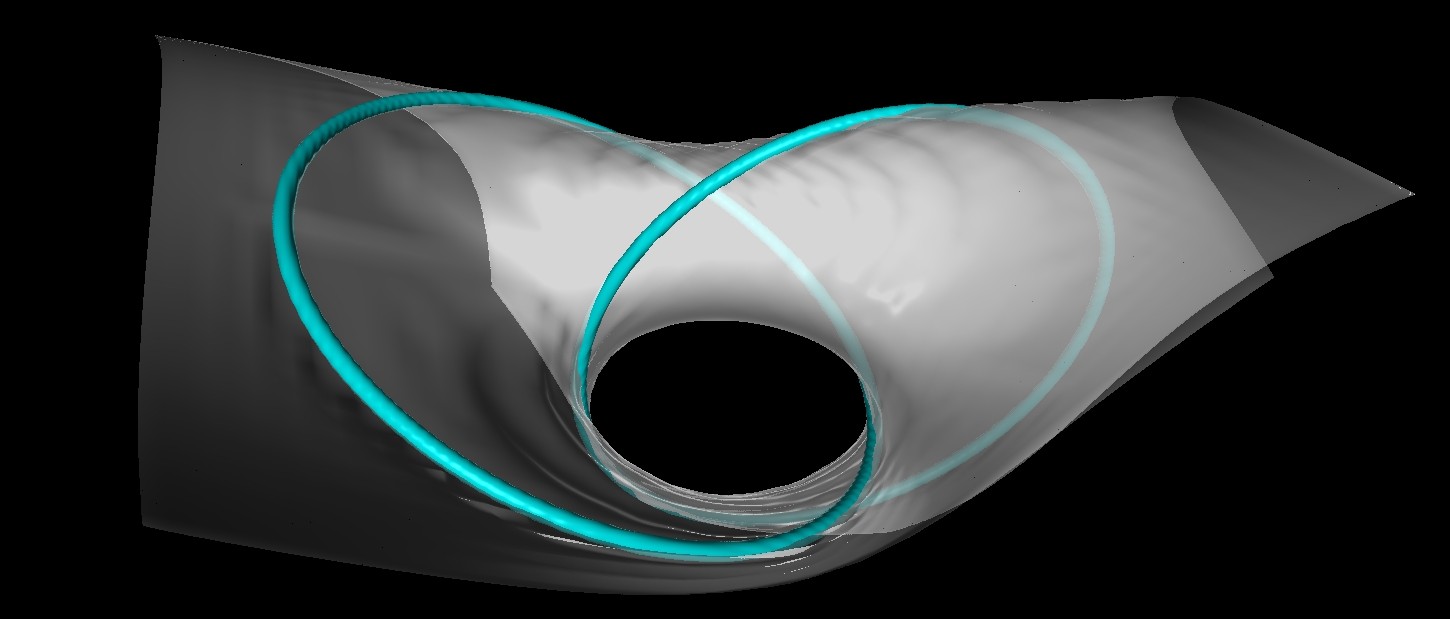}
    \vskip -1mm
    \includegraphics[width=8.5cm]{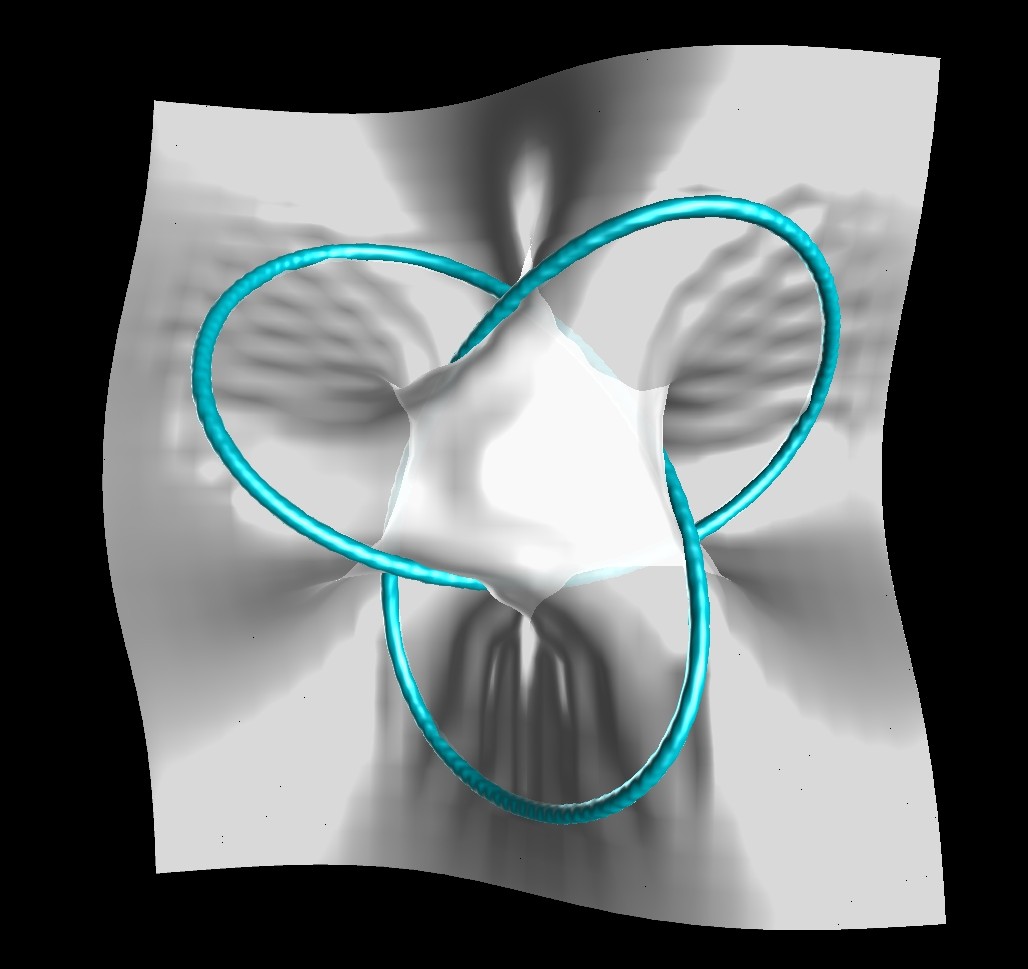}
    \caption{Renderings of the surface of zero phase for two knots in
      a quantum fluid. Top: two linked rings, note the surface has one
      hole. Bottom: trefoil knot, with three holes. The number of
      holes is associated to the number of turns the vector that lies
      on the surface perpendicular to the vortex lines does as it
      moves along the curve.}
    \label{ctephase}
\end{figure}


{\bf Helicity in quantum flows.} 
Low temperature quantum flows and BECs can be modeled as a field of
weakly interacting bosons of mass $m$ using the GPE,
\begin{align}
    i \hbar \frac{\partial \Psi}{\partial t} = - \frac{\hbar^2}{2m}
    \nabla^2 \Psi + g \vert \Psi \vert^2 \Psi,
    \label{gpe}
\end{align}
where $\Psi$ is the system's wavefunction and $g$ is proportional to the
scattering length. The flow matches the behavior of a classical, ideal,
and compressible potential fluid (i.e., it has no vorticity), except at
points where a topological singularity takes place. These topological
defects are the so-called quantum vortex lines where circulation is
quantized and given by $\Gamma = \oint_C {\bf v}(\ell) \,d\ell = 4 \pi
\alpha$, with ${\bf v}$ the flow velocity and $\alpha =
\hbar/(2m)$. The vorticity $\omega$ of the flow is thus given by
\begin{align}
    {\bm \omega} ({\bf r}) = \Gamma \int \d s \frac{\d {\bf r}'}{\d s}
    \delta^{(3)}({\bf r} - {\bf r}'(s)),
    \label{vortdelta}
\end{align}
where ${\bf r}(s)$ is the position of the vortex lines.

From the wavefunction, the particle density is given by
\begin{align}
    n=\overline{\Psi} \Psi,
\end{align}
and the velocity field can be obtained from
\begin{align}
    {\bf v} = \frac{\cal P} {n},
    \label{eq:vel}
\end{align}
where ${\cal P}$ is the unit mass momentum density
\begin{align}
    {\cal P}_j = 2 \alpha \frac {\overline{\Psi} \partial_j \Psi - \Psi
        \partial_j \overline{\Psi}} {2\, i} 
\end{align}
(notice that these definitions are analogous to those derived via the
Madelung transformation $\Psi = \sqrt{n} e^{i \phi}$, where the
velocity is given by ${\bf v} = 2 \alpha {\bm \nabla} \phi$). At a
distance $r\to0$ from a straight vortex line these quantities are
known \cite{Nore97a} to behave as $n\sim {r^2}$ and ${\bf v} = 2\alpha
{\bf e}_\theta/r$ where ${\bf e}_\theta$ is the azimuthal unit vector
and $r$ the radial distance in a cylindrical coordinate system $({\bf
  e}_r,{\bf e}_\theta,{\bf e}_z)$ having its origin on the straight
vortex line. Thus, the velocity ${\bf v}$ has an $r^{-1}$ singularity
{\it perpendicular} to the vortex line.

\begin{figure}
    \centering
    \includegraphics[width=8.5cm]{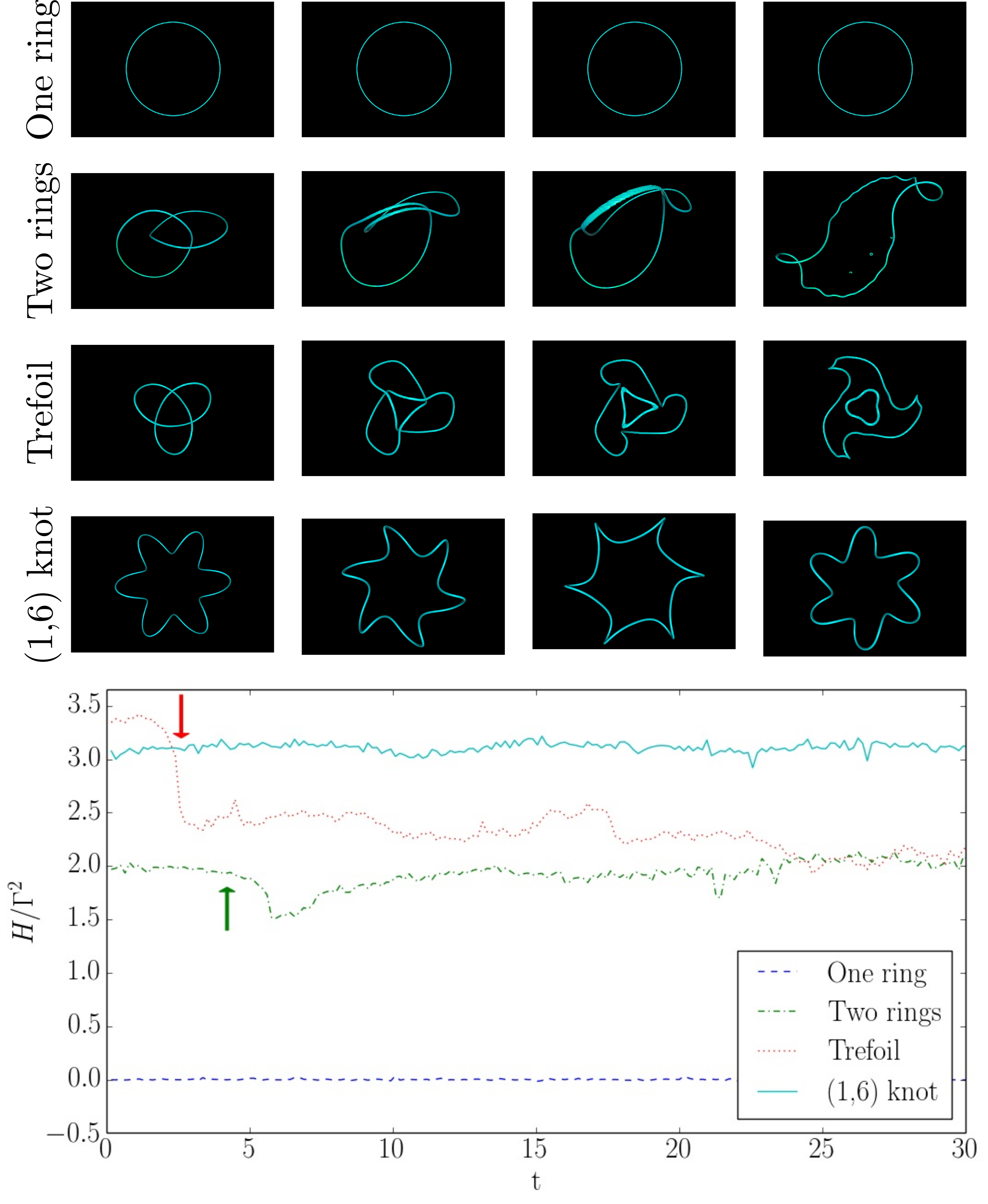}
    \caption{Time evolution of the helicity for four quantum vortex
      configurations. At the top, snapshots of the configurations
      at different times are shown. The single ring only moves at
      constant speed. The two rings and the trefoil reconnect at
      times marked by the vertical arrows. When reconnection takes
      place between two anti-parallel vortex lines (as in the two
      rings), helicity does not change. In the trefoil reconnection
      takes place simultaneously at three points and helicity changes
      abruptly at the time indicated by the red arrow; later it decays
      slowly to its final value. The (1,6)-torus knot deforms without
      reconnecting, and its helicity does not change.}
    \label{helknots}
\end{figure}

Therefore, as the vorticity (see Eq.(\ref{vortdelta})) also has a singularity {\it parallel} to those lines, the
standard definition of helicity
\begin{align}
    {\cal H}=\int \d{\bf r}\, {\bm \omega}({\bf r})\cdot {\bf v}({\bf r}),
    \label{ed:defhel}
\end{align}
is not well behaved, as it involves the product of two singular
distributions.  The idea of the {\it regularized} helicity is to replace
in Eq.~\eqref{ed:defhel} the field $\bf v$ by a regularized smooth field
${\bf v}_{\rm reg}$ having no divergences perpendicular to the line, and
the same regular behavior as ${\bf v}$ parallel to the line.

Starting from Eq.~\eqref{eq:vel}, we can regularize the velocity along
vortex lines (where $\Psi = 0$) by Taylor expanding $\Psi$ to first
order in the numerator and the denominator, arriving at
\begin{align*}
    v_\parallel=\frac {2 \alpha}{2 i} \frac {{\cal W}_j\left[(\partial_j
            \partial_l \Psi) \partial_l(\overline{\Psi}))-(\partial_j
            \partial_l \overline{\Psi}) \partial_l(\Psi))\right]}
    {\sqrt{{\cal W}_l{\cal W}_l}(\partial_m\Psi)(\partial_m
        \overline{\Psi})},
\end{align*}
where
\begin{align}
    {\cal W}_j = \epsilon_{jkl} \partial_k {\cal P}_l = \frac {2 \alpha}
    {i} \epsilon_{jkl}{\partial_k \overline{\Psi}  \partial_l \Psi}
    \label{rotrhov}
\end{align}
is a smooth field oriented along the vortex line. Then, we can define
the regularized helicity
\begin{align}
    {\cal H}=\int \d{\bf r}\, {\bm \omega}({\bf r})\cdot {\bf v}_{\rm reg}({\bf
        r}),
    \label{ed:defreg}
\end{align}
with ${\bf v}_{\rm reg}={v_\parallel {\cal W}}/{\sqrt{{\cal W}_j{\cal W}_j}}$.
We show next how this regularized helicity still holds the geometrical
interpretations valid for the standard one.

\begin{figure*}
    \centering
    \includegraphics[width=13cm]{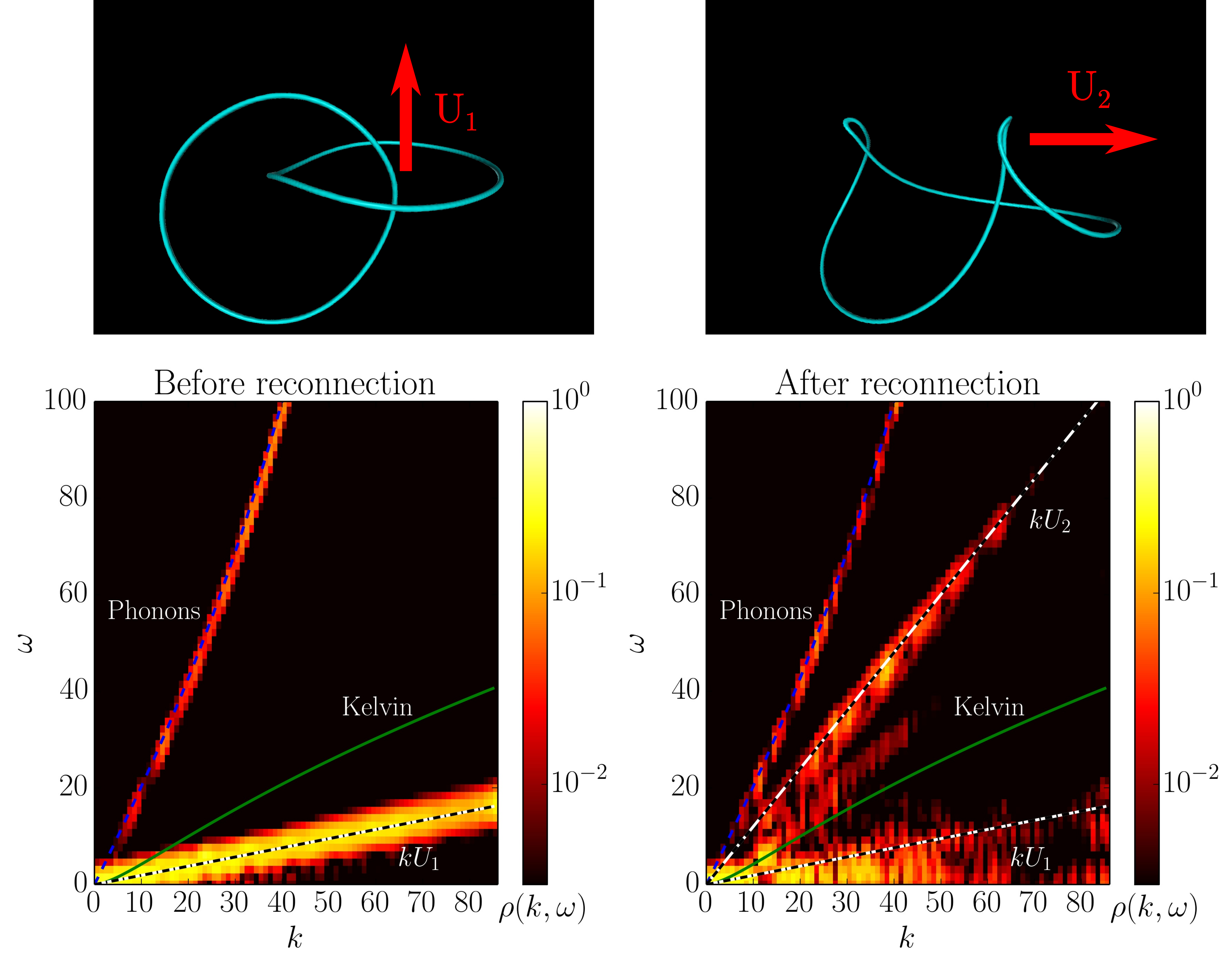}
    \caption{Spatiotemporal spectrum for the two rings before (left) 
        and after reconnection (right). The dashed blue line
        corresponds to the dispersion relation of sound waves, the
        solid green line to Kelvin waves, the dash-dotted line to
        sweeping with velocity $U_1$ (i.e., $\omega = U_1 k$), and the
        dash-triple dotted line to sweeping with $\omega = U_2
        k$. Sweeping concentrates most of the power, and only one
        energetic mode with $k\approx 11$ may be compatible with the
        dispersion relation of Kelvin waves.}
    \label{fig2}
\end{figure*}


{\bf Relation with writhe.} 
For an isolated structure, helicity can be decomposed into twist
(loosely speaking, the total number of helical turns a ribbon does),
and writhe (the ``coiling'' of the structure). Let's start by
analyzing the relation between the regularized helicity and the
writhe. For a single curve, the writhe $Wr$ is, by definition 
\cite{Klenin00}, given by the expression
\begin{align}
    Wr=\frac{1}{4 \pi} \frac{\int \int (\d{\bf r} \times \d{\bf r}_1)
        \cdot ({\bf r}-{\bf r}_1)} {|({\bf r}-{\bf r}_1)|^3} .
\end{align}

It is easy to see that if one uses a velocity field ${\bf V}({\bf r})$
given by the Biot-Savart law
\begin{align}
    {\bf V}({\bf r})=\frac{\Gamma}{4 \pi} \frac{\int \d{\bf r}_1 \times
        ({\bf r}-{\bf r}_1)} {|({\bf r}-{\bf r}_1)|^3},
\end{align}
where ${\bf r}_1$ corresponds to the position of the vortex lines, and
the vorticity as defined in Eq.~ \eqref{vortdelta}, then helicity
${\cal H}$ is given by 
\begin{align*}
    {\cal H} &= \int {\bf V}({\bf r})\cdot {\bm \omega}({\bf r}) \d V
                   = \Gamma \int {\bf V}({\bf r})\cdot \d{\bf r},
    \\
    &=\frac{\Gamma^2}{4 \pi} \frac{\int \int \d{\bf r}\cdot (\d{\bf r}_1
        \times ({\bf r}-{\bf r}_1))} {|({\bf r}-{\bf r}_1)|^3} .
\end{align*}
From the identity $({\bf a}\times {\bf b})\cdot {\bf c}={\bf
    a}\cdot({\bf b}\times {\bf c})$ one finds that in this simple case (for
a {\it single} line)
\begin{align*}
    {\cal H}=\Gamma^2 Wr .
\end{align*}


{\bf Regularized helicity defined as the twist of constant phase ribbon.}
First we recall that the twist  $Tw$ of a ribbon (defined by {\it
  both} a curve $\bf{r}(s)$, and a vector ${\bf U}(s)$ perpendicular
to the curve) is defined by the integral over the curve
\begin{align}
    Tw=\frac{1}{2 \pi} \int \left( \frac{d\bf{U}}{\d s}\times
        {\bf{U}}\right) \cdot
    \frac{d{\bf r}}{\d s} \d s  .
    \label{eq:twist}
\end{align}
One can further show that \cite{Moffatt92}
\begin{align}
    Tw=N + \frac{1}{2 \pi} \int \tau(s) \d s ,
\end{align}
where $\tau$ is the torsion, and $N$ the number of turns round the curve
of $\bf U$ in the Frenet-Serret frame (see {\it Methods}). The
regularized helicity can be presented in a purely geometrical
way. Under the GPE, constant phase surfaces will intersect on the
vortex lines. Now consider a line at a close distance of the vortex
line and lying on a constant phase surface (note that we could
construct an equivalent line in the classical Biot-Savart case by
requiring the line to be perpendicular to the velocity field). The
vortex line and the constant phase line define a ribbon. Now, using
Eqs.~\eqref{vortdelta}, \eqref{rotrhov} and \eqref{eq:twist} we can
see that
\begin{align*}
    {\cal H} = \Gamma^2 \, Tw .
\end{align*}
Note that, by construction, the circulation along the constant phase
line is zero.

As an illustration, Fig.~\ref{ctephase} shows renderings of surfaces of
zero phase for two knots in a quantum fluid. The presence of a hole
indicates that the vector perpendicular to the vortex line lying on this
surface does a whole turn as it moves along the vortex. Each of these
turns contributes by one quantum $\Gamma^2$ to the intrinsic twist,
and therefore to the helicity.


{\bf Knots in quantum flows.}
We consider four different initial conditions: one unknotted and
unlinked ring, two unknotted but linked rings, a trefoil knot, and a
(1,6)-torus knot (see {\it Methods} for details on the numerical scheme
used, and on the preparation of the initial conditions). Snapshots at
different times during their evolution are shown in
Fig.~\ref{helknots}. The single ring just moves at constant speed
(parallel to the axis of axisymmetry of the ring) without any 
deformation. The two linked rings deform and reconnect, the
reconnection taking place along segments (i.e., not at a single point)
of each ring, both of them aligning anti-parallely before
reconnecting. The trefoil knot undergoes three reconnections, which
happen simultaneously and, in contrast with the two linked rings, do
not involve long aligned segments. Finally, the (1,6)-torus knot deforms and 
moves but never reconnects. Videos showing the evolution of each
configuration can be found online \cite{Videos}.

The evolution of the regularized helicity (normalized by $\Gamma^2$)
for each configuration as a function of time is also shown in
Fig.~\ref{helknots}. A red arrow marks the time at which the trefoil
reconnects, and a green arrow the moment when the two linked rings
reconnect.

All four configurations start at the expected value of helicity. We
verified that this value is the same as the one obtained by other
methods to compute helicity in quantum flows (see, e.g.,
\cite{Zuccher15}). The single ring moves at constant speed without
deformations, and helicity remains constant at zero. 
The two rings move towards each
other, and align to reconnect two long anti-parallel segments (see the
third pannel of the snapshots). At that time there is a small drop of the
regularized helicity (associated with the fact that the regularization
is not well defined while the reconnection takes place), but then the
helicity remains constant around its original value of 2, even though
there remains only one ring after reconnection. This is to be expected
for anti-parallel reconnection, as predicted in \cite{Laing15}, and in
agreement with previous results \cite{Scheeler14}. As is clear from
the visualizations, the helicity in the link of the two rings gets
converted into a helical deformation (writhe) of the single ring. The
trefoil reconnects at three points simultaneously, and the vortex
lines are not anti-parallel at the moment of reconnection. As a
result, helicity rapidly drops by one quantum, from an initial value
of $\approx 3.4$. Remarkably, it then continues
dropping slowly until it reaches a new steady value of 2 quanta at $t
\approx 25$. As will be shown next, this decay is associated with the
excitation of helical waves along the two vortex rings resulting from
the reconnection. Finally, the (1,6)-torus knot deforms substantially as it
evolves, but its helicity remains around its initial value of 3 quanta.

Several conclusions can be drawn. First, there exist stable helical
solutions of the GPE where vortex knots do not reconnect. Second, in
agreement with previous studies \cite{Laing15,Zuccher15} antiparallel
reconnections conserve helicity. Third, and in disagreement with
previous claims, helicity is not conserved in all cases. Moreover,
helicity can vary rapidly during reconnection, or slowly afterwards by
a yet unclear mechanism. Below we show that this mechanism is the
emission of phonons by the non-linear interaction of helical
Kelvin-waves excited along the vortex lines.

\begin{figure*}
    \centering
    \includegraphics[width=13cm]{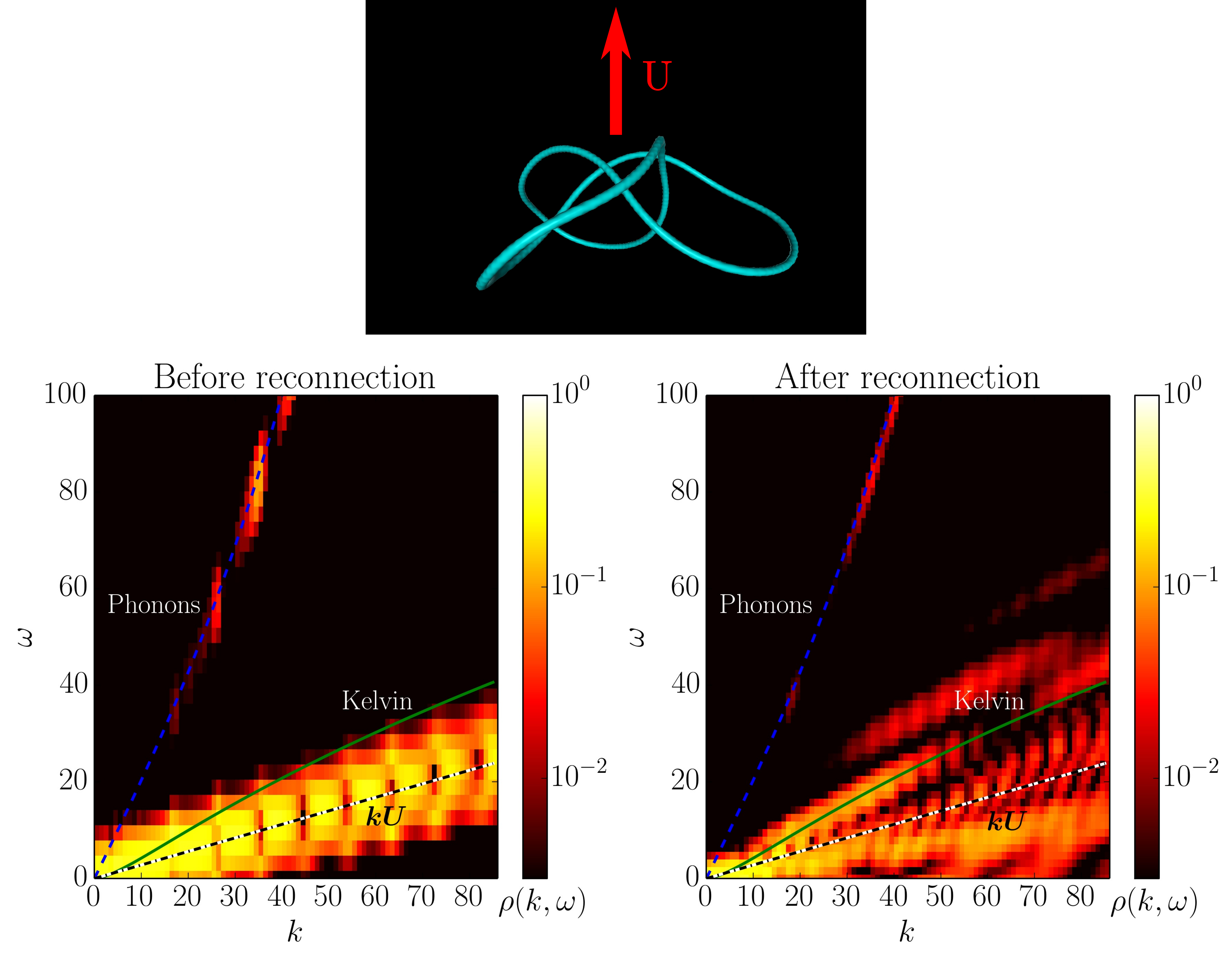}
    \caption{Spatiotemporal spectrum for the trefoil before (left) 
    and after the reconnection (right). The dashed blue line corresponds to
    sound waves, the solid green line to Kelvin waves, and the
    dash-dotted line to sweeping with $\omega = U k$. A broad range of
    modes compatible with the dispersion relation of Kelvin waves is
    excited after reconnection, and sound waves are visible at high
    frequencies.}
    \label{fig3}
\end{figure*}

{\bf Excitation of Kelvin waves by reconnection.} 
To understand the process that results in the slow depletion of
helicity, we compute the spatiotemporal spectrum of particle density
$\rho (k,\omega)$, before and after the reconnection, for the two rings
(Fig.~\ref{fig2}) and for the trefoil (Fig.~\ref{fig3}). This spectrum
is a useful tool to identify waves and flow displacements in complex
flows \cite{Clark15a,Clark15b}.  The GPE can sustain two types of
waves that will be of interest in the following: sound waves and
Kelvin waves. Sound waves follow the Bogoliubov dispersion relation
$\omega_B (k) = c k \sqrt{1 + \xi^2 k^2/2}$, where $c=\sqrt{g
  \vert\Psi\vert^2/m}$ is the speed of sound, and $\xi =
\sqrt{\hbar^2/(2mg \vert\Psi\vert^2)}$ is the coherence length
\cite{Nore97a}.  Kelvin waves follow the dispersion relation
\begin{align}
    \omega_K (k) = \frac{2 c \xi}{\sqrt{2} a^2} \left(
        1 \pm \sqrt{1 + \frac{K_0 (ka)}{K_1 (ka)}}
    \right),
\end{align}
where $a$ is the radius of the vortex core, and $K_0$ and $K_1$ are
modified Bessel functions.

\begin{figure}
    \centering
    \includegraphics[width=8.5cm]{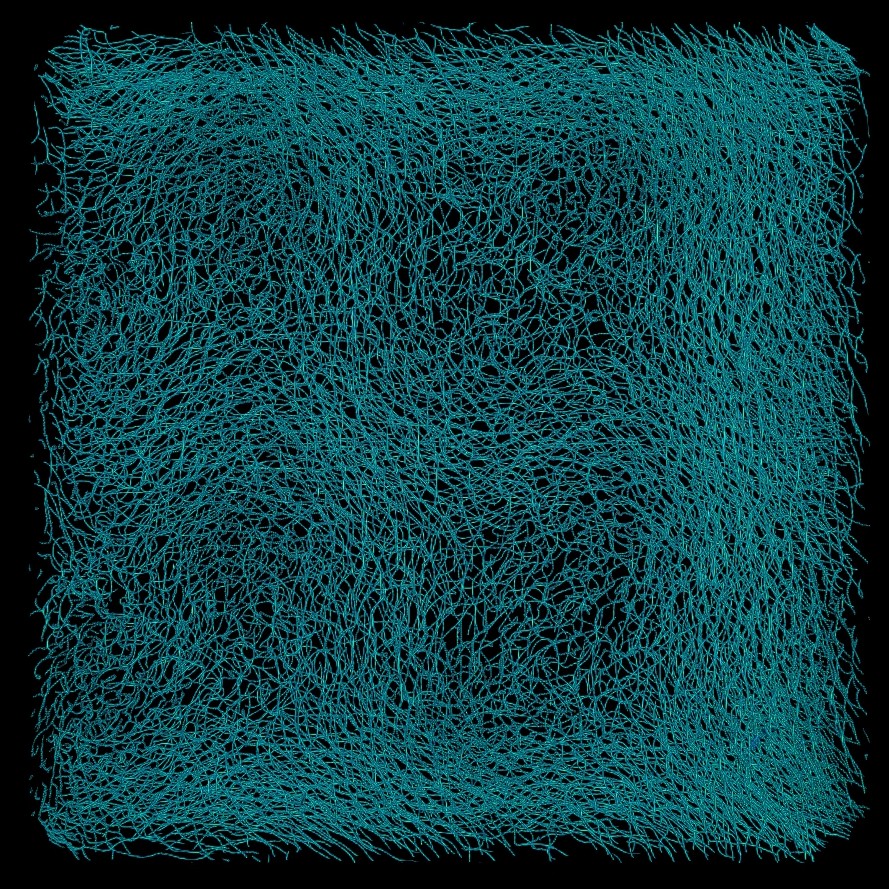}
    \caption{Rendering of vortex lines in a quantum flow with
      helicity. The regularized helicity is equal to $3$, matching the
      value expected for the classical flow at large scales. Normalizing
  by the circulation quanta, the helicity of this flow is 
  $\approx 480000\Gamma^2$.}
    \label{abc}
\end{figure}

In Fig.~\ref{fig2}, before reconnection takes place, the two rings
move towards each other at a mean velocity $U_1$. This appears in the
spatiotemporal spectrum as sweeping of the vortices, i.e., a
concentration of power near the region with $\omega = U_1 k$
(excitations corresponding to sound waves can also be
identified). After reconnection, the vortex still moves slowly with a
velocity close to $U_1$, but the reconnected points separate fast from
each other with velocity $U_2$. The sweeping of regions with low
density (the vortex lines) associated with this velocity is also
visible in the spectrum. There is almost no excitation of modes
compatible with Kelvin waves, except for a single mode with $k\approx
11$ which lies on top of the dispersion relation
$\omega_K(k=11)$. Indeed, in the last snapshot of the two rings in
Fig.~\ref{helknots}, a small helical perturbation with this wavenumber
can be observed (see also the movie in the supplemental material).

The spatiotemporal spectrum for the trefoil, for which helicity is not
conserved, is very different. Before reconnection the vortex knot moves
with mean velocity $U$, and sweeping with $\omega = U k$ can be observed
in the spectrum. After reconnection the motion of the two rings is
complex, although both structures still move with an average velocity
$U$ (a trace can be seen in the spectrum). However, the most remarkable
feature in the spectrum is the excitation of a broad range of modes
compatible with the dispersion relation of Kelvin waves, and the
excitation of sound waves only at high frequencies. The broad and
continuous range of Kelvin wave modes indicates the development of a
non-linear Kelvin wave cascade \cite{Clark15a}: as multiple modes are
excited, they can interact non-linearly and transfer their energy to
larger wavenumbers, where the energy in the modes (and their helicity)
can be dissipated by phonon emission \cite{Vinen03}. Indeed, the Kelvin
waves fade away once helicity reaches its new steady state value of
$\approx 2$ quanta in Fig.~\ref{helknots}.

{\bf Helicity in complex quantum flows.}
Finally, we show the regularized helicity is robust even for quantum
turbulence, where hundreds of thousands of knots can be
present in the flow. Figure \ref{abc} shows the three-dimensional
rendering of a helical flow, with a distribution of vortex lines such
that the flow large-scale structure corresponds to the superposition
of two classical ABC flows at wavenumbers $k=1$ and at $k=2$ (see {\it
  Methods}). The flow was computed in a grid of $2048^3$ grid points,
with a very high density of vortex lines. Computation of the
regularized helicity over the quantum flow gives the expected value of
3, matching the classical value. In units of $\Gamma^2$ this value
corresponds to $\approx 480000$ links.

Our observations allow the study of helicity in complex quantum knots
and helical quantum turbulence, and can indicate in the future new
links between helicity, Kelvin waves, and excitation of waves after
reconnection, important for other areas such as fluid dynamics and
space physics. The recent experimental creation of quantum knots in 
the laboratory \cite{Hall16} shows also a promising application for
the quantification and evolution of the topological complexity of
quantum vortices.

{\bf Acknowledgments}
The authors acknowledge financial support from Grant No. ECOS-Sud
A13E01, and from computing hours in the CURIE supercomputer granted by
Project TGCC-GENCI No. x20152a7493.

\bibliography{ms}

\begin{thebibliography}{34}%
\makeatletter
\providecommand \@ifxundefined [1]{%
 \@ifx{#1\undefined}
}%
\providecommand \@ifnum [1]{%
 \ifnum #1\expandafter \@firstoftwo
 \else \expandafter \@secondoftwo
 \fi
}%
\providecommand \@ifx [1]{%
 \ifx #1\expandafter \@firstoftwo
 \else \expandafter \@secondoftwo
 \fi
}%
\providecommand \natexlab [1]{#1}%
\providecommand \enquote  [1]{``#1''}%
\providecommand \bibnamefont  [1]{#1}%
\providecommand \bibfnamefont [1]{#1}%
\providecommand \citenamefont [1]{#1}%
\providecommand \href@noop [0]{\@secondoftwo}%
\providecommand \href [0]{\begingroup \@sanitize@url \@href}%
\providecommand \@href[1]{\@@startlink{#1}\@@href}%
\providecommand \@@href[1]{\endgroup#1\@@endlink}%
\providecommand \@sanitize@url [0]{\catcode `\\12\catcode `\$12\catcode
  `\&12\catcode `\#12\catcode `\^12\catcode `\_12\catcode `\%12\relax}%
\providecommand \@@startlink[1]{}%
\providecommand \@@endlink[0]{}%
\providecommand \url  [0]{\begingroup\@sanitize@url \@url }%
\providecommand \@url [1]{\endgroup\@href {#1}{\urlprefix }}%
\providecommand \urlprefix  [0]{URL }%
\providecommand \Eprint [0]{\href }%
\providecommand \doibase [0]{http://dx.doi.org/}%
\providecommand \selectlanguage [0]{\@gobble}%
\providecommand \bibinfo  [0]{\@secondoftwo}%
\providecommand \bibfield  [0]{\@secondoftwo}%
\providecommand \translation [1]{[#1]}%
\providecommand \BibitemOpen [0]{}%
\providecommand \bibitemStop [0]{}%
\providecommand \bibitemNoStop [0]{.\EOS\space}%
\providecommand \EOS [0]{\spacefactor3000\relax}%
\providecommand \BibitemShut  [1]{\csname bibitem#1\endcsname}%
\let\auto@bib@innerbib\@empty
\bibitem [{\citenamefont {Brandenburg}\ and\ \citenamefont
  {Subramanian}(2005)}]{Brandenburg05}%
  \BibitemOpen
  \bibfield  {author} {\bibinfo {author} {\bibfnamefont {A.}~\bibnamefont
  {Brandenburg}}\ and\ \bibinfo {author} {\bibfnamefont {K.}~\bibnamefont
  {Subramanian}},\ }\href {\doibase 10.1016/j.physrep.2005.06.005} {\bibfield
  {journal} {\bibinfo  {journal} {Physics Reports}\ }\textbf {\bibinfo {volume}
  {417}},\ \bibinfo {pages} {1} (\bibinfo {year} {2005})}\BibitemShut {NoStop}%
\bibitem [{\citenamefont {Rasmussen}\ and\ \citenamefont
  {Blanchard}(1998)}]{Rasmussen98}%
  \BibitemOpen
  \bibfield  {author} {\bibinfo {author} {\bibfnamefont {E.~N.}\ \bibnamefont
  {Rasmussen}}\ and\ \bibinfo {author} {\bibfnamefont {D.~O.}\ \bibnamefont
  {Blanchard}},\ }\href {\doibase
  10.1175/1520-0434(1998)013<1148:ABCOSD>2.0.CO;2} {\bibfield  {journal}
  {\bibinfo  {journal} {Weather and Forecasting}\ }\textbf {\bibinfo {volume}
  {13}},\ \bibinfo {pages} {1148} (\bibinfo {year} {1998})}\BibitemShut
  {NoStop}%
\bibitem [{\citenamefont {Rorai}\ \emph {et~al.}(2013)\citenamefont {Rorai},
  \citenamefont {Rosenberg}, \citenamefont {Pouquet},\ and\ \citenamefont
  {Mininni}}]{Rorai13}%
  \BibitemOpen
  \bibfield  {author} {\bibinfo {author} {\bibfnamefont {C.}~\bibnamefont
  {Rorai}}, \bibinfo {author} {\bibfnamefont {D.}~\bibnamefont {Rosenberg}},
  \bibinfo {author} {\bibfnamefont {A.}~\bibnamefont {Pouquet}}, \ and\
  \bibinfo {author} {\bibfnamefont {P.~D.}\ \bibnamefont {Mininni}},\ }\href
  {\doibase 10.1103/PhysRevE.87.063007} {\bibfield  {journal} {\bibinfo
  {journal} {Phys.\ Rev.\ E}\ }\textbf {\bibinfo {volume} {87}},\ \bibinfo
  {pages} {063007} (\bibinfo {year} {2013})}\BibitemShut {NoStop}%
\bibitem [{\citenamefont {Marino}\ \emph {et~al.}(2013)\citenamefont {Marino},
  \citenamefont {Mininni}, \citenamefont {Rosenberg},\ and\ \citenamefont
  {Pouquet}}]{Marino13}%
  \BibitemOpen
  \bibfield  {author} {\bibinfo {author} {\bibfnamefont {R.}~\bibnamefont
  {Marino}}, \bibinfo {author} {\bibfnamefont {P.~D.}\ \bibnamefont {Mininni}},
  \bibinfo {author} {\bibfnamefont {D.}~\bibnamefont {Rosenberg}}, \ and\
  \bibinfo {author} {\bibfnamefont {A.}~\bibnamefont {Pouquet}},\ }\href
  {\doibase 10.1103/PhysRevE.87.033016} {\bibfield  {journal} {\bibinfo
  {journal} {Phys.\ Rev.\ E}\ }\textbf {\bibinfo {volume} {87}} (\bibinfo
  {year} {2013}),\ 10.1103/PhysRevE.87.033016}\BibitemShut {NoStop}%
\bibitem [{\citenamefont {Moffatt}(1969)}]{Moffatt69}%
  \BibitemOpen
  \bibfield  {author} {\bibinfo {author} {\bibfnamefont {H.~K.}\ \bibnamefont
  {Moffatt}},\ }\href {\doibase 10.1017/S0022112069000991} {\bibfield
  {journal} {\bibinfo  {journal} {J.\ Fluid Mech.}\ }\textbf {\bibinfo {volume}
  {35}},\ \bibinfo {pages} {117} (\bibinfo {year} {1969})}\BibitemShut
  {NoStop}%
\bibitem [{\citenamefont {Moffatt}\ and\ \citenamefont
  {Ricca}(1992)}]{Moffatt92}%
  \BibitemOpen
  \bibfield  {author} {\bibinfo {author} {\bibfnamefont {H.~K.}\ \bibnamefont
  {Moffatt}}\ and\ \bibinfo {author} {\bibfnamefont {R.~L.}\ \bibnamefont
  {Ricca}},\ }\href {\doibase 10.1098/rspa.1992.0159} {\bibfield  {journal}
  {\bibinfo  {journal} {Proceedings of the Royal Society of London A:
  Mathematical, Physical and Engineering Sciences}\ }\textbf {\bibinfo {volume}
  {439}},\ \bibinfo {pages} {411} (\bibinfo {year} {1992})}\BibitemShut
  {NoStop}%
\bibitem [{\citenamefont {Dennis}\ and\ \citenamefont
  {Hannay}(2005)}]{Dennis05}%
  \BibitemOpen
  \bibfield  {author} {\bibinfo {author} {\bibfnamefont {M.~R.}\ \bibnamefont
  {Dennis}}\ and\ \bibinfo {author} {\bibfnamefont {J.~H.}\ \bibnamefont
  {Hannay}},\ }\href {\doibase 10.1098/rspa.2005.1527} {\bibfield  {journal}
  {\bibinfo  {journal} {Proceedings of the Royal Society of London A:
  Mathematical, Physical and Engineering Sciences}\ }\textbf {\bibinfo {volume}
  {461}},\ \bibinfo {pages} {3245} (\bibinfo {year} {2005})}\BibitemShut
  {NoStop}%
\bibitem [{\citenamefont {Ricca}\ and\ \citenamefont {Berger}(2008)}]{Ricca08}%
  \BibitemOpen
  \bibfield  {author} {\bibinfo {author} {\bibfnamefont {R.~L.}\ \bibnamefont
  {Ricca}}\ and\ \bibinfo {author} {\bibfnamefont {M.~A.}\ \bibnamefont
  {Berger}},\ }\href {\doibase 10.1063/1.881574} {\bibfield  {journal}
  {\bibinfo  {journal} {Physics Today}\ }\textbf {\bibinfo {volume} {49}},\
  \bibinfo {pages} {28} (\bibinfo {year} {2008})}\BibitemShut {NoStop}%
\bibitem [{\citenamefont {Moffatt}(2014)}]{Moffatt14}%
  \BibitemOpen
  \bibfield  {author} {\bibinfo {author} {\bibfnamefont {H.~K.}\ \bibnamefont
  {Moffatt}},\ }\href {\doibase 10.1073/pnas.1400277111} {\bibfield  {journal}
  {\bibinfo  {journal} {Proc.\ Natl.\ Acad.\ Sci.\ U.S.A.}\ }\textbf {\bibinfo
  {volume} {111}},\ \bibinfo {pages} {3663} (\bibinfo {year}
  {2014})}\BibitemShut {NoStop}%
\bibitem [{\citenamefont {Vologodskii}\ \emph {et~al.}(1998)\citenamefont
  {Vologodskii}, \citenamefont {Crisona}, \citenamefont {Laurie}, \citenamefont
  {Pieranski}, \citenamefont {Katritch}, \citenamefont {Dubochet},\ and\
  \citenamefont {Stasiak}}]{Vologodskii98}%
  \BibitemOpen
  \bibfield  {author} {\bibinfo {author} {\bibfnamefont {A.~V.}\ \bibnamefont
  {Vologodskii}}, \bibinfo {author} {\bibfnamefont {N.~J.}\ \bibnamefont
  {Crisona}}, \bibinfo {author} {\bibfnamefont {B.}~\bibnamefont {Laurie}},
  \bibinfo {author} {\bibfnamefont {P.}~\bibnamefont {Pieranski}}, \bibinfo
  {author} {\bibfnamefont {V.}~\bibnamefont {Katritch}}, \bibinfo {author}
  {\bibfnamefont {J.}~\bibnamefont {Dubochet}}, \ and\ \bibinfo {author}
  {\bibfnamefont {A.}~\bibnamefont {Stasiak}},\ }\href {\doibase
  10.1006/jmbi.1998.1696} {\bibfield  {journal} {\bibinfo  {journal} {Journal
  of Molecular Biology}\ }\textbf {\bibinfo {volume} {278}},\ \bibinfo {pages}
  {1} (\bibinfo {year} {1998})}\BibitemShut {NoStop}%
\bibitem [{\citenamefont {Dennis}\ \emph {et~al.}(2010)\citenamefont {Dennis},
  \citenamefont {King}, \citenamefont {Jack}, \citenamefont {O'Holleran},\ and\
  \citenamefont {Padgett}}]{Dennis10}%
  \BibitemOpen
  \bibfield  {author} {\bibinfo {author} {\bibfnamefont {M.~R.}\ \bibnamefont
  {Dennis}}, \bibinfo {author} {\bibfnamefont {R.~P.}\ \bibnamefont {King}},
  \bibinfo {author} {\bibfnamefont {B.}~\bibnamefont {Jack}}, \bibinfo {author}
  {\bibfnamefont {K.}~\bibnamefont {O'Holleran}}, \ and\ \bibinfo {author}
  {\bibfnamefont {M.~J.}\ \bibnamefont {Padgett}},\ }\href {\doibase
  10.1038/nphys1504} {\bibfield  {journal} {\bibinfo  {journal} {Nature
  Physics}\ }\textbf {\bibinfo {volume} {6}},\ \bibinfo {pages} {118} (\bibinfo
  {year} {2010})}\BibitemShut {NoStop}%
\bibitem [{\citenamefont {Kedia}\ \emph {et~al.}(2013)\citenamefont {Kedia},
  \citenamefont {Bialynicki-Birula}, \citenamefont {Peralta-Salas},\ and\
  \citenamefont {Irvine}}]{Kedia13}%
  \BibitemOpen
  \bibfield  {author} {\bibinfo {author} {\bibfnamefont {H.}~\bibnamefont
  {Kedia}}, \bibinfo {author} {\bibfnamefont {I.}~\bibnamefont
  {Bialynicki-Birula}}, \bibinfo {author} {\bibfnamefont {D.}~\bibnamefont
  {Peralta-Salas}}, \ and\ \bibinfo {author} {\bibfnamefont {W.~T.~M.}\
  \bibnamefont {Irvine}},\ }\href {\doibase 10.1103/PhysRevLett.111.150404}
  {\bibfield  {journal} {\bibinfo  {journal} {Phys.\ Rev.\ Lett.}\ }\textbf
  {\bibinfo {volume} {111}},\ \bibinfo {pages} {150404} (\bibinfo {year}
  {2013})}\BibitemShut {NoStop}%
\bibitem [{\citenamefont {Hussain}\ and\ \citenamefont
  {Duraisamy}(2011)}]{Hussain11}%
  \BibitemOpen
  \bibfield  {author} {\bibinfo {author} {\bibfnamefont {F.}~\bibnamefont
  {Hussain}}\ and\ \bibinfo {author} {\bibfnamefont {K.}~\bibnamefont
  {Duraisamy}},\ }\href {\doibase 10.1063/1.3532039} {\bibfield  {journal}
  {\bibinfo  {journal} {Phys.\ Fluids}\ }\textbf {\bibinfo {volume} {23}},\
  \bibinfo {pages} {021701} (\bibinfo {year} {2011})}\BibitemShut {NoStop}%
\bibitem [{\citenamefont {Kleckner}\ and\ \citenamefont
  {Irvine}(2013)}]{Kleckner13}%
  \BibitemOpen
  \bibfield  {author} {\bibinfo {author} {\bibfnamefont {D.}~\bibnamefont
  {Kleckner}}\ and\ \bibinfo {author} {\bibfnamefont {W.~T.~M.}\ \bibnamefont
  {Irvine}},\ }\href {\doibase 10.1038/nphys2560} {\bibfield  {journal}
  {\bibinfo  {journal} {Nature Phys.}\ }\textbf {\bibinfo {volume} {9}},\
  \bibinfo {pages} {253} (\bibinfo {year} {2013})}\BibitemShut {NoStop}%
\bibitem [{\citenamefont {Bewley}\ \emph {et~al.}(2008)\citenamefont {Bewley},
  \citenamefont {Paoletti}, \citenamefont {Sreenivasan},\ and\ \citenamefont
  {Lathrop}}]{Bewley08}%
  \BibitemOpen
  \bibfield  {author} {\bibinfo {author} {\bibfnamefont {G.~P.}\ \bibnamefont
  {Bewley}}, \bibinfo {author} {\bibfnamefont {M.~S.}\ \bibnamefont
  {Paoletti}}, \bibinfo {author} {\bibfnamefont {K.~R.}\ \bibnamefont
  {Sreenivasan}}, \ and\ \bibinfo {author} {\bibfnamefont {D.~P.}\ \bibnamefont
  {Lathrop}},\ }\href {\doibase 10.1073/pnas.0806002105} {\bibfield  {journal}
  {\bibinfo  {journal} {Proc.\ Natl.\ Acad.\ Sci.\ U.S.A.}\ } (\bibinfo {year}
  {2008}),\ 10.1073/pnas.0806002105}\BibitemShut {NoStop}%
\bibitem [{\citenamefont {Zuccher}\ \emph {et~al.}(2012)\citenamefont
  {Zuccher}, \citenamefont {Caliari}, \citenamefont {Baggaley},\ and\
  \citenamefont {Barenghi}}]{Zuccher12}%
  \BibitemOpen
  \bibfield  {author} {\bibinfo {author} {\bibfnamefont {S.}~\bibnamefont
  {Zuccher}}, \bibinfo {author} {\bibfnamefont {M.}~\bibnamefont {Caliari}},
  \bibinfo {author} {\bibfnamefont {A.~W.}\ \bibnamefont {Baggaley}}, \ and\
  \bibinfo {author} {\bibfnamefont {C.~F.}\ \bibnamefont {Barenghi}},\ }\href
  {\doibase 10.1063/1.4772198} {\bibfield  {journal} {\bibinfo  {journal}
  {Phys.\ Fluids}\ }\textbf {\bibinfo {volume} {24}},\ \bibinfo {pages}
  {125108} (\bibinfo {year} {2012})}\BibitemShut {NoStop}%
\bibitem [{\citenamefont {Scheeler}\ \emph {et~al.}(2014)\citenamefont
  {Scheeler}, \citenamefont {Kleckner}, \citenamefont {Proment}, \citenamefont
  {Kindlmann},\ and\ \citenamefont {Irvine}}]{Scheeler14}%
  \BibitemOpen
  \bibfield  {author} {\bibinfo {author} {\bibfnamefont {M.~W.}\ \bibnamefont
  {Scheeler}}, \bibinfo {author} {\bibfnamefont {D.}~\bibnamefont {Kleckner}},
  \bibinfo {author} {\bibfnamefont {D.}~\bibnamefont {Proment}}, \bibinfo
  {author} {\bibfnamefont {G.~L.}\ \bibnamefont {Kindlmann}}, \ and\ \bibinfo
  {author} {\bibfnamefont {W.~T.~M.}\ \bibnamefont {Irvine}},\ }\href {\doibase
  10.1073/pnas.1407232111} {\bibfield  {journal} {\bibinfo  {journal} {Proc.\
  Natl.\ Acad.\ Sci.\ U.S.A.}\ }\textbf {\bibinfo {volume} {111}},\ \bibinfo
  {pages} {15350} (\bibinfo {year} {2014})}\BibitemShut {NoStop}%
\bibitem [{\citenamefont {Laing}\ \emph {et~al.}(2015)\citenamefont {Laing},
  \citenamefont {Ricca},\ and\ \citenamefont {Sumners}}]{Laing15}%
  \BibitemOpen
  \bibfield  {author} {\bibinfo {author} {\bibfnamefont {C.~E.}\ \bibnamefont
  {Laing}}, \bibinfo {author} {\bibfnamefont {R.~L.}\ \bibnamefont {Ricca}}, \
  and\ \bibinfo {author} {\bibfnamefont {D.~W.~L.}\ \bibnamefont {Sumners}},\
  }\href {\doibase 10.1038/srep09224} {\bibfield  {journal} {\bibinfo
  {journal} {Scientific Reports}\ }\textbf {\bibinfo {volume} {5}},\ \bibinfo
  {pages} {9224} (\bibinfo {year} {2015})}\BibitemShut {NoStop}%
\bibitem [{\citenamefont {Zuccher}\ and\ \citenamefont
  {Ricca}(2015)}]{Zuccher15}%
  \BibitemOpen
  \bibfield  {author} {\bibinfo {author} {\bibfnamefont {S.}~\bibnamefont
  {Zuccher}}\ and\ \bibinfo {author} {\bibfnamefont {R.~L.}\ \bibnamefont
  {Ricca}},\ }\href {\doibase 10.1103/PhysRevE.92.061001} {\bibfield  {journal}
  {\bibinfo  {journal} {Physical Review E}\ }\textbf {\bibinfo {volume} {92}},\
  \bibinfo {pages} {061001} (\bibinfo {year} {2015})}\BibitemShut {NoStop}%
\bibitem [{\citenamefont {Kimura}\ and\ \citenamefont
  {Moffatt}(2014)}]{Kimura14}%
  \BibitemOpen
  \bibfield  {author} {\bibinfo {author} {\bibfnamefont {Y.}~\bibnamefont
  {Kimura}}\ and\ \bibinfo {author} {\bibfnamefont {H.~K.}\ \bibnamefont
  {Moffatt}},\ }\href {\doibase 10.1017/jfm.2014.233} {\bibfield  {journal}
  {\bibinfo  {journal} {J.\ Fluid Mech.}\ }\textbf {\bibinfo {volume} {751}},\
  \bibinfo {pages} {329} (\bibinfo {year} {2014})}\BibitemShut {NoStop}%
\bibitem [{\citenamefont {Hall}\ \emph {et~al.}(2016)\citenamefont {Hall},
  \citenamefont {Ray}, \citenamefont {Tiurev}, \citenamefont {Ruokokoski},
  \citenamefont {Gheorghe},\ and\ \citenamefont {Mottonen}}]{Hall16}%
  \BibitemOpen
  \bibfield  {author} {\bibinfo {author} {\bibfnamefont {D.~S.}\ \bibnamefont
  {Hall}}, \bibinfo {author} {\bibfnamefont {M.~W.}\ \bibnamefont {Ray}},
  \bibinfo {author} {\bibfnamefont {K.}~\bibnamefont {Tiurev}}, \bibinfo
  {author} {\bibfnamefont {E.}~\bibnamefont {Ruokokoski}}, \bibinfo {author}
  {\bibfnamefont {A.~H.}\ \bibnamefont {Gheorghe}}, \ and\ \bibinfo {author}
  {\bibfnamefont {M.}~\bibnamefont {Mottonen}},\ }\href
  {http://dx.doi.org/10.1038/nphys3624} {\bibfield  {journal} {\bibinfo
  {journal} {Nat Phys}\ }\textbf {\bibinfo {volume} {advance online
  publication}},\  (\bibinfo {year} {2016})}\BibitemShut {NoStop}%
\bibitem [{\citenamefont {Fonda}\ \emph {et~al.}(2014)\citenamefont {Fonda},
  \citenamefont {Meichle}, \citenamefont {Ouellette}, \citenamefont {Hormoz},\
  and\ \citenamefont {Lathrop}}]{Fonda14}%
  \BibitemOpen
  \bibfield  {author} {\bibinfo {author} {\bibfnamefont {E.}~\bibnamefont
  {Fonda}}, \bibinfo {author} {\bibfnamefont {D.~P.}\ \bibnamefont {Meichle}},
  \bibinfo {author} {\bibfnamefont {N.~T.}\ \bibnamefont {Ouellette}}, \bibinfo
  {author} {\bibfnamefont {S.}~\bibnamefont {Hormoz}}, \ and\ \bibinfo {author}
  {\bibfnamefont {D.~P.}\ \bibnamefont {Lathrop}},\ }\href {\doibase
  10.1073/pnas.1312536110} {\bibfield  {journal} {\bibinfo  {journal} {Proc.\
  Natl.\ Acad.\ Sci.\ U.S.A.}\ }\textbf {\bibinfo {volume} {111}},\ \bibinfo
  {pages} {4707} (\bibinfo {year} {2014})}\BibitemShut {NoStop}%
\bibitem [{\citenamefont {Thomson}(1880)}]{Thomson80}%
  \BibitemOpen
  \bibfield  {author} {\bibinfo {author} {\bibfnamefont {W.}~\bibnamefont
  {Thomson}},\ }\href {\doibase 10.1080/14786448008626912} {\bibfield
  {journal} {\bibinfo  {journal} {Philos. Mag.}\ }\textbf {\bibinfo {volume}
  {10}},\ \bibinfo {pages} {155} (\bibinfo {year} {1880})}\BibitemShut
  {NoStop}%
\bibitem [{\citenamefont {Kozik}\ and\ \citenamefont
  {Svistunov}(2004)}]{Kozik04}%
  \BibitemOpen
  \bibfield  {author} {\bibinfo {author} {\bibfnamefont {E.}~\bibnamefont
  {Kozik}}\ and\ \bibinfo {author} {\bibfnamefont {B.}~\bibnamefont
  {Svistunov}},\ }\href {\doibase 10.1103/PhysRevLett.92.035301} {\bibfield
  {journal} {\bibinfo  {journal} {Phys.\ Rev.\ Lett.}\ }\textbf {\bibinfo
  {volume} {92}},\ \bibinfo {pages} {035301} (\bibinfo {year}
  {2004})}\BibitemShut {NoStop}%
\bibitem [{\citenamefont {L'vov}\ and\ \citenamefont
  {Nazarenko}(2010)}]{Lvov10}%
  \BibitemOpen
  \bibfield  {author} {\bibinfo {author} {\bibfnamefont {V.~S.}\ \bibnamefont
  {L'vov}}\ and\ \bibinfo {author} {\bibfnamefont {S.}~\bibnamefont
  {Nazarenko}},\ }\href {\doibase 10.1134/S002136401008014X} {\bibfield
  {journal} {\bibinfo  {journal} {J.\ Exp.\ Theor.\ Phys.\ Lett.}\ }\textbf
  {\bibinfo {volume} {91}},\ \bibinfo {pages} {428} (\bibinfo {year}
  {2010})}\BibitemShut {NoStop}%
\bibitem [{\citenamefont {Clark~di Leoni}\ \emph {et~al.}(2015)\citenamefont
  {Clark~di Leoni}, \citenamefont {Mininni},\ and\ \citenamefont
  {Brachet}}]{Clark15a}%
  \BibitemOpen
  \bibfield  {author} {\bibinfo {author} {\bibfnamefont {P.}~\bibnamefont
  {Clark~di Leoni}}, \bibinfo {author} {\bibfnamefont {P.~D.}\ \bibnamefont
  {Mininni}}, \ and\ \bibinfo {author} {\bibfnamefont {M.~E.}\ \bibnamefont
  {Brachet}},\ }\href {\doibase 10.1103/PhysRevA.92.063632} {\bibfield
  {journal} {\bibinfo  {journal} {Phys. Rev. A}\ }\textbf {\bibinfo {volume}
  {92}},\ \bibinfo {pages} {063632} (\bibinfo {year} {2015})}\BibitemShut
  {NoStop}%
\bibitem [{\citenamefont {Nore}\ \emph {et~al.}(1997)\citenamefont {Nore},
  \citenamefont {Abid},\ and\ \citenamefont {Brachet}}]{Nore97a}%
  \BibitemOpen
  \bibfield  {author} {\bibinfo {author} {\bibfnamefont {C.}~\bibnamefont
  {Nore}}, \bibinfo {author} {\bibfnamefont {M.}~\bibnamefont {Abid}}, \ and\
  \bibinfo {author} {\bibfnamefont {M.~E.}\ \bibnamefont {Brachet}},\ }\href
  {\doibase 10.1063/1.869473} {\bibfield  {journal} {\bibinfo  {journal}
  {Phys.\ Fluids}\ }\textbf {\bibinfo {volume} {9}},\ \bibinfo {pages} {2644}
  (\bibinfo {year} {1997})}\BibitemShut {NoStop}%
\bibitem [{\citenamefont {Klenin}\ and\ \citenamefont
  {Langowski}(2000)}]{Klenin00}%
  \BibitemOpen
  \bibfield  {author} {\bibinfo {author} {\bibfnamefont {K.}~\bibnamefont
  {Klenin}}\ and\ \bibinfo {author} {\bibfnamefont {J.}~\bibnamefont
  {Langowski}},\ }\href {\doibase
  10.1002/1097-0282(20001015)54:5<307::AID-BIP20>3.0.CO;2-Y} {\bibfield
  {journal} {\bibinfo  {journal} {Biopolymers}\ }\textbf {\bibinfo {volume}
  {54}},\ \bibinfo {pages} {307} (\bibinfo {year} {2000})}\BibitemShut
  {NoStop}%
\bibitem [{Vid()}]{Videos}%
  \BibitemOpen
  \href@noop {} {\bibinfo  {journal}
  {\url{http://wp.df.uba.ar/mininni/movies/#quantum}}\ }\BibitemShut {NoStop}%
\bibitem [{\citenamefont {di~Leoni}\ \emph {et~al.}(2015)\citenamefont
  {di~Leoni}, \citenamefont {Cobelli},\ and\ \citenamefont
  {Mininni}}]{Clark15b}%
  \BibitemOpen
\bibfield  {journal} {  }\bibfield  {author} {\bibinfo {author} {\bibfnamefont
  {P.~C.}\ \bibnamefont {di~Leoni}}, \bibinfo {author} {\bibfnamefont
  {P.}~\bibnamefont {Cobelli}}, \ and\ \bibinfo {author} {\bibfnamefont
  {P.}~\bibnamefont {Mininni}},\ }\href@noop {} {\bibfield  {journal} {\bibinfo
   {journal} {The European Physical Journal E}\ }\textbf {\bibinfo {volume}
  {38}},\ \bibinfo {pages} {1} (\bibinfo {year} {2015})}\BibitemShut {NoStop}%
\bibitem [{\citenamefont {Vinen}\ \emph {et~al.}(2003)\citenamefont {Vinen},
  \citenamefont {Tsubota},\ and\ \citenamefont {Mitani}}]{Vinen03}%
  \BibitemOpen
  \bibfield  {author} {\bibinfo {author} {\bibfnamefont {W.}~\bibnamefont
  {Vinen}}, \bibinfo {author} {\bibfnamefont {M.}~\bibnamefont {Tsubota}}, \
  and\ \bibinfo {author} {\bibfnamefont {A.}~\bibnamefont {Mitani}},\ }\href
  {\doibase 10.1103/PhysRevLett.91.135301} {\bibfield  {journal} {\bibinfo
  {journal} {Phys.\ Rev.\ Lett.}\ }\textbf {\bibinfo {volume} {91}},\ \bibinfo
  {pages} {135301} (\bibinfo {year} {2003})}\BibitemShut {NoStop}%
\bibitem [{\citenamefont {G\'omez}\ \emph
  {et~al.}(2005{\natexlab{a}})\citenamefont {G\'omez}, \citenamefont
  {Mininni},\ and\ \citenamefont {Dmitruk}}]{Gomez05a}%
  \BibitemOpen
  \bibfield  {author} {\bibinfo {author} {\bibfnamefont {D.~O.}\ \bibnamefont
  {G\'omez}}, \bibinfo {author} {\bibfnamefont {P.~D.}\ \bibnamefont
  {Mininni}}, \ and\ \bibinfo {author} {\bibfnamefont {P.}~\bibnamefont
  {Dmitruk}},\ }\href {\doibase 10.1016/j.asr.2005.02.099} {\bibfield
  {journal} {\bibinfo  {journal} {Advances in Space Research}\ }\textbf
  {\bibinfo {volume} {35}},\ \bibinfo {pages} {899} (\bibinfo {year}
  {2005}{\natexlab{a}})}\BibitemShut {NoStop}%
\bibitem [{\citenamefont {G\'omez}\ \emph
  {et~al.}(2005{\natexlab{b}})\citenamefont {G\'omez}, \citenamefont
  {Mininni},\ and\ \citenamefont {Dmitruk}}]{Gomez05b}%
  \BibitemOpen
  \bibfield  {author} {\bibinfo {author} {\bibfnamefont {D.~O.}\ \bibnamefont
  {G\'omez}}, \bibinfo {author} {\bibfnamefont {P.~D.}\ \bibnamefont
  {Mininni}}, \ and\ \bibinfo {author} {\bibfnamefont {P.}~\bibnamefont
  {Dmitruk}},\ }\href {\doibase 10.1238/Physica.Topical.116a00123} {\bibfield
  {journal} {\bibinfo  {journal} {Phys. Scripta}\ }\textbf {\bibinfo {volume}
  {2005}},\ \bibinfo {pages} {123} (\bibinfo {year}
  {2005}{\natexlab{b}})}\BibitemShut {NoStop}%
\bibitem [{\citenamefont {Mininni}\ \emph {et~al.}(2011)\citenamefont
  {Mininni}, \citenamefont {Rosenberg}, \citenamefont {Reddy},\ and\
  \citenamefont {Pouquet}}]{Mininni11}%
  \BibitemOpen
  \bibfield  {author} {\bibinfo {author} {\bibfnamefont {P.}~\bibnamefont
  {Mininni}}, \bibinfo {author} {\bibfnamefont {D.}~\bibnamefont {Rosenberg}},
  \bibinfo {author} {\bibfnamefont {R.}~\bibnamefont {Reddy}}, \ and\ \bibinfo
  {author} {\bibfnamefont {A.}~\bibnamefont {Pouquet}},\ }\href {\doibase
  10.1016/j.parco.2011.05.004} {\bibfield  {journal} {\bibinfo  {journal}
  {Parallel Computing}\ }\textbf {\bibinfo {volume} {37}},\ \bibinfo {pages}
  {316} (\bibinfo {year} {2011})}\BibitemShut {NoStop}%
\end{thebibliography}%

\newpage

\section*{Methods} 

{\bf Numerical scheme.}
The equations were integrated using GHOST 
\cite{Gomez05a,Gomez05b,Mininni11}, a three dimensional code which
uses a pseudospectral scheme with periodic boundary conditions to
compute spatial derivatives  and a fourth order Runge-Kutta scheme to
compute time derivatives. The ``2/3 rule'' is used for
de-aliasing. The code is parallelized using both MPI and OpenMP. The
vortex knots simulations were done using $256^3$ grid points, while
the ABC simulation was done with $2048^3$ grid points.

\vskip 0.5cm
{\bf Preparation method for knots initial data.}
The initial data preparation method is based on the one presented in
\cite{Scheeler14}. The method consists in calculating the velocity
field generated by a vortex line (or lines) ${\bf r}(s)$, which is then
integrated to get the phase of the wavefunction. The density at each
point in space is then calculated by using a Pad\'e approximation. One of
the two differences with the method presented in \cite{Scheeler14} is
that after doing this we first use the generated wavefunction as an
initial condition of the advected Real Guinzbug Landau equation, whose
stationary solutions are solutions of the GPE with minimal acoustic
energy \cite{Nore97a}, and then feed that solutions to the GPE, thereby
minimizing errors (specially those stemming from the Pad\'e
approximation). The other key difference is that our fields are
truly periodic. Instead of using an array of replicas to generate an
almost periodic field, we work in the Fourier domain using the Fourier
transform of the vorticity \eqref{vortdelta}, which as we evaluate only
at integer wavenumbers gives a perfectly periodic field. The velocity
field is then obtained by applying the inverse of the curl operator
(i.e., the Biot-Savart law).

\vskip 0.5cm
{\bf Preparation method for quantum ABC flow.}
The so-called ABC (Arnold, Beltrami and Childress) velocity field is a
maximal helicity stationary solution of Euler equations in which the
vorticity is parallel to the velocity, explicitly given by
\begin{align}
    {\bf u}_{\rm ABC}(x,y,z) &=& \left\{ \left[B
    \cos(k y) +
        C \sin(k z) \right] \hat{x} + \right. {} \nonumber \\
    && {} + \left[A \sin(k x) + C \cos(k z) \right] \hat{y} +
       {} \nonumber \\
    && {} + \left. \left[A \cos(k x) + B \sin(k y) \right]
       \hat{z} \right\}.
    \label{eq:ABC}
\end{align}
This velocity is the sum of three simple ($A=B=0$, $A=C=0$ and $B=C=0$)
flows. We first construct an ARGLE initial wavefunction for each of
these flows, and then take their product and run ARGLE. It is easy to
see that the $A=B=0$ flow is a constant $z$-dependent advection in each
$x-y$ slice. By Madelung's transformation the constant advection  
$C (\sin(k z) \hat{x}+ \cos(k z) \hat{y})$ should correspond to a
wavefunction
\begin{align}
    \Psi(x,y,z)=e^{i \frac {C \sin(k z)}{2 \alpha} x+\frac {C \cos(k
            z)}{2 \alpha} y}
\end{align}

In order to have a $2 \pi$-periodic initial data we initially set
\begin{align}
    \Psi(x,y,z)=e^{i [\frac {C \sin(k z)}{2 \alpha}] x+[\frac {C \cos(k
            z)}{2 \alpha}] y}
\end{align}
where $[a]$ stands for the integer nearer to $a$. 

The general initial data is made out of a product of such functions,
corresponding to non-zero $A$ and $B$ and various values of the
wavenumber $k$.  Note that the frustration (the relative difference between $C
    \sin(k z)/(2 \alpha)$ and an integer) goes down when $\alpha=c
    \xi/\sqrt{2} \to 0$).

\vskip 0.5cm
{\bf Frenet-Serret frame and equations.}
We recall that, given a 3D curve ${\bf r}(s)$, with ${\bf r}=(x,y,z)$
and $ds=\sqrt{dx^2+dy^2+dz^2}$, the standard Frenet-Serret tangent $\bf
T$, normal $\bf N$, and binormal $\bf B$ vectors are defined as
\begin{align}
    \frac{d\bf{r}}{ds} & =\bf{T} , \\
    \frac{d\bf{T}}{ds} \Big/ {\left\Vert{\frac{d\bf{T}}{ds}}\right\Vert}  &=\bf{N} , \\
    \bf{T} \times \bf{N}  &=\bf{B} .
\end{align}

These obey the Frenet-Serret equations
\begin{align}
    \frac{d\bf{T}}{ds}  &  = \kappa \bf{N} , \\
    \frac{d\bf{N}}{ds}  &  =-\kappa\bf{T+}\tau\bf{B} , \\
    \frac{d\bf{B}}{ds}  &  =-\tau\bf{N} . \\
\end{align}
where $\kappa$ is the curvature and $\tau$ the torsion.
\end{document}